\documentclass[prb,twocolumn,showpacs,amsmath,amssymb]{revtex4}
\usepackage{graphicx}
\usepackage{enumerate}
\newcommand \beq{\begin{eqnarray}}
\newcommand \eeq{\end{eqnarray}}

\newcommand{\bp}{\mbox{\boldmath $p$}}

\newcommand{\bk}{\mbox{\boldmath $k$}}

\begin{document}

\title{Chiral Gap and Collective Excitations in Monolayer Graphene\\
from Strong Coupling Expansion of Lattice Gauge Theory}
\author{Yasufumi Araki and Tetsuo Hatsuda}
\affiliation{
Department of Physics, The University of Tokyo, Tokyo 113-0033, Japan
}

\begin{abstract}
 Using the strong coupling expansion  of the compact and non-compact U(1) lattice gauge
 theories for monolayer graphene, we show analytically that 
 fermion bandgap and pseudo Nambu--Goldstone exciton ($\pi$-exciton)
 are dynamically generated due to chiral symmetry breaking. 
 The mechanism is similar to the generation of quark mass
 and pion excitation in quantum chromodynamics (QCD). We 
 derive a formula for the $\pi$-exciton analogous to the 
 Gell-Mann--Oakes--Renner (GOR) relation in QCD. Experimental
 confirmation of the GOR relation on a suspended monolayer
 graphene would be a clear evidence of chiral symmetry breaking. 
\end{abstract}
\maketitle

Graphene is a monoatomic layer of carbon atoms with a honeycomb lattice structure
 \cite{novoselov_2004,castro_neto_2009}.
  A novel  feature of graphene
   is that  electrons and holes
   at low energy have a  linear dispersion relation around
    two independent ``Dirac points'' in the momentum space  \cite{wallace_1947}.
 Then the charge carriers on graphene can be described by 
 massless Dirac quasiparticles \cite{Semenoff:1984dq}.
  The system has however a critical difference from relativistic electrons\cite{wallace_1947}; 
  the Fermi velocity $v_{_F}$ of the electrons on graphene is about 300 times smaller than 
  the speed of light $c$.  This leads to an effective
  enhancement of the  Coulomb interaction among Dirac quasiparticles.
    In such a  strong-coupling situation,  electrons and holes on graphene
   may form an exciton condensate and create a gap in the fermion 
  spectrum leading to  semimetal-insulator transition. 
  To show unambiguously that undoped monolayer graphene suspended in vacuum
   becomes an insulator  is one of the important theoretical challenges  \cite{CN09}.  
  Also, this problem has much in common with the dynamical breaking of 
      chiral symmetry in strongly-coupled relativistic field theories
      such as quantum chromodynamics (QCD) \cite{Hatsuda:1994pi}.  
 So far, various theoretical methods  such as the  
 Schwinger--Dyson equation \cite{khveshchenko_2001,gorbar_2002,gorbar_2009}, renormalization group equations
 \cite{son_2007} and lattice Monte Carlo methods \cite{hands_2008,drut_2009} 
 have been applied to study the dynamical  formation of the fermion gap 
 in low-energy effective theories of graphene. 

 The purpose of this Rapid Communication is to shed lights on the
  strong coupling regime of graphene at zero temperature 
   from an analytic  method of  strong coupling expansion
 (See Refs.\onlinecite{Nishida:2003uj,Miura:2009nu} for its recent
  applications in QCD).
  We start with a ``braneworld'' or ``reduced QED'' 
   model of graphene \cite{gorbar_2002,son_2007}
 in which (2+1)-dimensional Dirac fermions are 
  coupled to (3+1)-dimensional Coulomb field.
  After discretizing this model on a square lattice with
   compact U(1) gauge field and staggered fermion 
   \cite{hands_2008,drut_2009}, we carry out an expansion  
  by the inverse  Coulomb coupling
     and derive an effective action for the fermions.
       The exciton condensate (chiral condensate)
  and  the fermion gap (chiral gap) are obtained analytically
  from the resultant effective action.
  Properties of the pseudo Nambu--Goldstone (NG) excitation
 associated with the exciton condensation are also studied: In particular,
  we derive a mass formula  
 similar to the  Gell-Mann--Oakes--Renner relation (GOR) in QCD \cite{gmor}.

Low-energy Euclidean action for Dirac 
quasiparticles on a graphene interacting with the U(1) gauge field
is described by \cite{gorbar_2002,son_2007}
\begin{eqnarray}
S_E &=& \sum_{f}
 \int dx^{(3)} \ \bar{\psi}_f \left( D[A_4] +m \right) \psi_f \nonumber \\
 & &  + \frac{1}{2g^2} \sum_{j=1,2,3}  \int dx^{(4)} \ (\partial_j A_4)^2 ,
 \label{eq:effaction}
\end{eqnarray}
where the natural unit ($\hbar=c=1$) is taken.
Since this is an effective theory for  
 quasiparticles in the first Brillouin zone
of  the original honeycomb lattice, it has an intrinsic momentum cutoff
 $p_{_{\Lambda}} \lesssim \pi/a_{\rm hc}$ with  $a_{\rm hc}=1.42\ \text{\AA}$ being 
 the honeycomb lattice spacing.

The three-dimensional and four-dimensional Euclidean coordinates are denoted by
 $x^{(3)}=(\tau,x_1,x_2)$ and $x^{(4)}=(\tau,x_1,x_2,x_3)$, respectively.   
The Dirac spinor  $\psi_f$ has four components corresponding to 
2 (the number of sublattices)$\times$ 2 (the number of Dirac points).
The ``flavor index'' $f$ runs from $1$ up to the number of flavors $N$.
In this Rapid Communication we specifically focus on $N=2$ case,
which corresponds to the monolayer graphene with the up and down spins of electrons.
  The Dirac operator is defined as
  $D[A_4]= \gamma_4(\partial_4+iA_4) + v_{_F} (\gamma_1 \partial_1 + \gamma_2 \partial_2) $,
  where
$A_4$ is a temporal component of the gauge field. The gauge coupling constant
for the suspended graphene is $g^2=e^2/\epsilon_0$ with $e$ being the electric charge and 
 $\epsilon_0$ being the vacuum permittivity.
$g^2$ is reduced by the factor $2/(1+\varepsilon)$ on a substrate
with $\varepsilon$ being the dielectric constant of the substrate \cite{son_2007}.
 The Hermitian $\gamma$ matrices obey the standard relation 
  $\{\gamma_\mu,\gamma_\nu\}=2\delta_{\mu\nu}$.
The Fermi velocity reads
$v_{_F}=(3/2)t a_{\rm hc}=3.02\times 10^{-3}$ in the unit of light velocity, with
the hopping parameter $t \simeq 2.8\ \text{eV}$, \cite{reich}
obtained from the spectral slope observed on substrate.
 The bare mass $m$ corresponds to an explicit bandgap which may be 
  formed artificially on epitaxially grown graphene on substrate
   \cite{zhou_2007}
  or on graphene nanoribbon and nanomesh \cite{bai_2010}.

 Due to the small Fermi velocity, electron interactions are dominated by
 the Coulomb interaction, so that
   the spatial components of the gauge field, $A_{j=1,2,3}$,  can be neglected.
  With scaled variables, $\tau \rightarrow   \tau /v_{_F}$
  and $A_4 \rightarrow v_{_F} A_4$,  Dirac particles 
  have an effective mass $m_*=m/v_{_F}$ and an effective coupling $g_*^2 =g^2/v_{_F}$
   which is about 300 times larger than the 
   Coulomb coupling strength in the vacuum.
 In the chiral limit ($m\rightarrow 0$),
 Eq.(\ref{eq:effaction}) is invariant under  U(4) chiral transformation
  with 16 generators \cite{chiral-symmetry}: 
  $(1, \vec{\sigma}) \otimes (1,  \gamma_3, \gamma_5, \gamma_3 \gamma_5)$
   with $\gamma_5=\gamma_4 \gamma_1 \gamma_2 \gamma_3$. 
 Absence of $\gamma_3$ in $D[A_4]$ is the reason for such 
 large chiral symmetry.
 
 A regularized form of Eq.(\ref{eq:effaction}) on a hypothetical 
 square lattice with a lattice spacing $a=\pi/p_{_{\Lambda}}$ reads  \cite{drut_2009}
\begin{eqnarray}
\! \! \! \! \! S_F&=& \sum_{x^{(3)}}  \left[  \frac{1}{2} \sum_{\mu=1,2,4}
 \left( V_{\mu}^+(x)-V_{\mu}^-(x) \right)  + m_{*} M(x) \right] ,
\label{eq:latticeaction-F} \\
\! \! \! \! \!  S_G&=& \frac{1}{g_*^2} \sum_{x^{(4)}} \sum_{j=1,2,3}
\left[1-{\rm Re} \left( U_4(x) U_4^{\dagger}(x+\hat{j}) \right)\right].
 \label{eq:latticeaction-G}  
\end{eqnarray}
 Here all the dimensionful quantities are scaled by  $a$.
The U(1) gauge action $S_G$ is  written in terms of a time-like 
 link variable $U_4(x) = \exp( i \theta(x) )$  
  with $-\pi < \theta \le \pi$.
 The fermionic action $S_F$ is written in terms of the staggered fermion $\chi$ through
\begin{eqnarray}
 M(x)&=& \sum_b \bar{\chi}_b(x)\chi_b(x),  \\
 V_{\mu}^+(x)&=& \sum_b \eta_{\mu}(x)\bar{\chi}_b(x)U_{\mu}(x) \chi_b(x+\hat{\mu}), \nonumber \\
 V_{\mu}^-(x)&=& \sum_b \eta_{\mu}(x)\bar{\chi}_b(x+\hat{\mu}) U_{\mu}^{\dagger}(x)   \chi_b(x),
\end{eqnarray}
 with  $\mu =  1, 2, 4$ and   $U_{1,2}(x)=1$. 
$b$ is the staggered flavor index which runs from $1$ to $N/2$,
since $2^3$ doublers emerging from one staggered fermion on a 3-dimensional square lattice
can be identified with four Dirac 
components times two ``flavors''\cite{hands_2008,drut_2009}.
The monolayer graphene corresponds to $N=2$.
The staggered phase factors  $\eta_{\mu}$ 
  are
$ \eta_4(x)=1, \eta_1(x)=(-1)^{\tau},  \eta_2(x)=(-1)^{\tau+x_1}$, 
 and $\eta_3(x)=(-1)^{\tau+x_1+x_2} \equiv \epsilon(x) $.

In the chiral limit,
 fermion action for each flavor in $S_F$ is invariant under  
 U(1)$_{_{\rm V}}$ $\times$U(1)$_{_{\rm A}}$
  chiral transformations;  $(\chi_b(x), \bar{\chi}_b(x))
\xrightarrow[ ]{\rm V} 
(e^{i\xi_{_{\rm V}}} \chi_b(x), e^{-i\xi_{_{\rm V}}}\bar{\chi}_b(x))$ 
and  $(\chi_b(x), \bar{\chi}_b(x))
  \xrightarrow[ ]{\rm A} (e^{i\xi_{_{\rm A}} \epsilon(x)} \chi_b(x), 
  e^{i\xi_{_{\rm A}} \epsilon(x)} \bar{\chi}_b(x) )$.
  These are remnants
   of global U(4) chiral symmetry of Eq.(\ref{eq:effaction}). \cite{Kogut:1982ds}
 Under  the U(1)$_{_{\rm A}}$ rotation, we have
 $M(x) \rightarrow e^{2i\xi_{_{\rm A}} \epsilon(x)}M(x)$ 
 and $V_{\mu}^{\pm}(x) \rightarrow V_{\mu}^{\pm}(x)$, so that
  the   chiral condensate $\langle \bar{\chi} \chi \rangle$ serves as an order   
  parameter for  the spontaneous symmetry breaking,
  U(1)$_{_{\rm V}}$$\times$U(1)$_{_{\rm A}}$$\rightarrow$U(1)$_{_{\rm V}}$.

\begin{figure}[tb]
\begin{center}
\includegraphics[width=6.5cm]{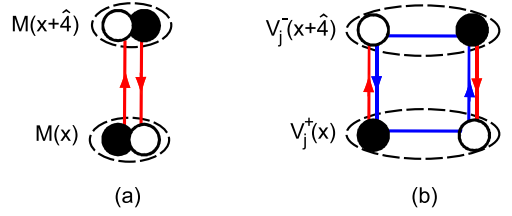}
\end{center}
\vspace{-0.5cm}
\caption{Induced four-fermion interaction in the strong coupling
 expansion.
The open (filled)  circle represent
 $\chi$ ($\bar{\chi}$).
(a) In the LO,  the time-like links (red arrows) 
in  $S_F$ 
 cancel with each other to leave a spatially local interaction.
(b) In the NLO,
 the time-link links in $S_F$ are canceled by those in 
 $S_G$  (blue arrows)  to leave a spatially non-local interaction.
}
\label{fig:links}
\end{figure}

We define an expansion parameter, 
$\beta \equiv 1/g_{*}^2$,
 so that the strong coupling limit corresponds to $\beta \rightarrow 0$.
 (By taking $v_{_F}\sim 0.003$, $\beta$ in the vacuum is estimated as 0.04,
  while that on the SiO$_2$ substrate is 0.1.) 
 Then, the partition function expanded by $S_G \sim O(\beta)$  becomes 
\begin{equation}
 Z \! = \!\! \int \! [d\chi d\bar{\chi}][d\theta] \! \left[ \sum_{n=0}^{\infty}
\! \frac{(-S_G)^n }{n!} e^{-S_F} \right] \!\! = \!\! \int
 [d\chi d\bar{\chi}] e^{-S_{\chi}}. \!\!\!
 \label{eq:part-Z}
\end{equation}
The $\theta$  integration can be analytically performed
 order by order in $\beta$ for general $N$\cite{Drouffe:1983fv}: 
 When the link variables $e^{i\theta}$ cancel with each other,
 the  fermion self-interactions up to $2N$-fermi term are induced.
Hereafter, we will focus only on the monolayer case ($N=2$).\cite{NG4}
 Up to $O(\beta)$, we obtain
\begin{eqnarray}
\label{eq:4-fermi-free}
& & \! \! \! \! \! \! \! \!   \! \! \!  
S_{\chi}= \sum_{x^{(3)}}  \left[ \frac{1}{2} \sum_{j=1,2} 
  \left( V_j^+(x)-V_j^-(x) \right)  + m_{*} M(x) \right]  
\nonumber \\
\label{eq:4-fermi-0}
& & \! \! \! \! \!  
 -\frac{1}{4} \sum_{x^{(3)}} M(x) M(x+\hat{4}) \nonumber \\
\label{eq:4-fermi-1}
& & \! \! \! \! \!  + \frac{\beta}{8} \sum_{x^{(3)}} \sum_{j=1,2}
\left( V_j^+(x) V_j^-(x+\hat{4}) + (V_j^+ \leftrightarrow V_j^-) \right)  . 
\end{eqnarray}
The second line in Eq.(\ref{eq:4-fermi-0}) is the  leading-order (LO) term 
of  $O(\beta^0)$ which is 
 local (non-local) in space (time)  as shown in Fig.\ref{fig:links}(a).
 The third line is the next-to-leading-order (NLO) term of 
  $O(\beta)$  which is non-local in both space and time
 as shown in Fig.\ref{fig:links}(b).  
 Note that the gauge field propagating along the third spatial
 dimension $x_3$ starts to appear  at $O(\beta^3)$ in the strong coupling expansion. 
  The non-local four-fermi interactions in Eq.(\ref{eq:4-fermi-0})
  can be liniarized 
   by  the extended Stratonovich--Hubbard transformation \cite{Miura:2009nu}:
 $\exp(\alpha AB) \sim  
 \int [d\varphi d\varphi^*] \exp[-\alpha(|\varphi|^2-A\varphi-B\varphi^*)]$, 
 where $A$ and $B$ are fermion bilinears and  $\alpha$ is a positive constant. 
 By introducing two complex auxiliary fields $\phi(x)$ and $\lambda(x)$ 
 corresponding to the LO and NLO terms and 
 integrating out the fermion fields, we arrive at
 $Z= \int [d\phi d\phi^*][d\lambda d\lambda^*]e^{-S_{\rm eff}(\phi,\lambda)}$,
 where the axial U(1)$_{_{\rm A}}$ rotation induces the transformation,
  $\phi(x) \rightarrow e^{-2i \xi_{_{\rm A}} \epsilon(x)} \phi(x)$ 
 and $\lambda (x) \rightarrow \lambda (x) $.

  In the mean-field approximation where
  fluctuations of  $\phi$ and $\lambda$ are neglected, 
   free energy per unit space-time lattice cell at zero temperature,
    $F_\text{eff}(\phi)$, can be
    obtained after  eliminating  $\lambda$ by using the stationary condition 
  ($\delta S_{\rm eff}(\phi,\lambda)/\delta \lambda=0$):
 \begin{eqnarray}
F_\text{eff}(\phi)
&=& \frac{1}{4} {|\phi|^2}  - \frac{1}{2} \int_k  \ln \left[ G^{-1}(\bk;{\phi}) \right] 
\nonumber \\
& &\! \! \! \! \! \! \! \! \! \! \! \! \! \!  - \frac{\beta}{4} \sum_{j=1,2} 
\left[ \int_k  G(\bk;{\phi}) \sin^2 k_j \right]^2 + O(\beta^2).
\label{eq:effpot-NLO}
\end{eqnarray} 
Here $G^{-1}(\bk;{\phi})
=\sum_{j=1,2} \sin^2 k_j + |m_{*} - \phi/2|^2$ is the 
two dimensional bosonic propagator with an effective mass,
 $m_{*} - \phi/2$ and  
 $\int_{k} \equiv  
  \frac{1}{\pi^2} \int_{-\pi/2}^{\pi/2} dk_1 \int_{-\pi/2}^{\pi/2} dk_2$.
  Alternative way to derive Eq.(\ref{eq:effpot-NLO})
  is to treat the $O(\beta)$ term of  
  Eq.(\ref{eq:4-fermi-1}) as a first order perturbation.
 
\begin{figure}[tb]
\begin{center}
\includegraphics[width=6.5cm]{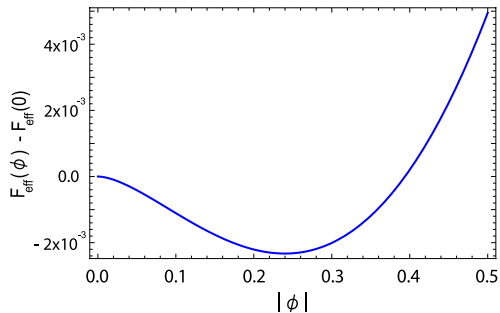}
\end{center}
\vspace{-0.5cm}
\caption{The free energy $F_\text{eff}(\phi)$ in the lattice unit 
as a function of $|\phi|$  for $\beta=0$ and $m=0$.}
\label{fig:NLO}
\end{figure}

  The free energy $F_{\rm eff}(\phi)$ 
  in the chiral and  srtong coupling limit ($m=0$, $\beta=0$)  
 is shown in Fig.\ref{fig:NLO}  for illustration.
  From Eq.(\ref{eq:effpot-NLO}), we find that
    $F_{\rm eff}(\phi \rightarrow \infty) \sim |\phi|^2$ 
  due to the tree-level term, while
  $F_{\rm eff}(\phi\rightarrow 0) \sim {\rm const.} + |\phi|^2 \ln |\phi|^2$  
  due to the fermion one-loop term. Therefore,
  we can exactly show, in the case when $N=2$, that
  dynamical  chiral symmetry breaking takes place in the 
  strong coupling limit.
  Since the $O(\beta)$ correction from the third term in
  Eq.(\ref{eq:effpot-NLO}) grows as $|\phi|$ increases,
  the chiral condensate, 
  $\sigma \equiv | \langle \bar{\chi} \chi \rangle |$, 
  is a decreasing function of $\beta$.  Up to
   the linear terms in $\beta$ and $m$, we have
\begin{eqnarray}
\label{eq:sigma-exp}
\sigma \simeq (0.240 - 0.297 \beta + 0.0239\ ma) a^{-2},
\end{eqnarray}
where we recover the lattice spacing $a$.
If we employ 
 $a^{-1} \sim a^{-1}_{\rm hc} = 1.39 \ {\rm keV}$ as 
 a typical cutoff scale of our effective theory,
 we obtain
 $\sigma \simeq 
 \left[\left(0.680 - 0.421 \beta + \frac{1.39\ m}{\rm eV} \right)
 {\rm keV} \right]^2$.
Note that our approach is limited to the strong coupling regime, so that
 it is inappropriate to  extract 
   the critical coupling $\beta_{\rm c}$ for semimetal--insulator
 transition from Eq.(\ref{eq:sigma-exp}).
  The total fermion mass $M_F$ is a sum of the dynamical mass and the bare mass
 in $G(\bk;\phi)$, 
\begin{equation}
M_F \equiv 
({\sigma a^2}/{2}) ({v_{_F}}/{a}) +m, 
\end{equation}
which reduces to $(0.523 - 0.623 \beta ) \ {\rm eV} \ + 3.05 m$
for $a \sim a_{\rm hc}$.
 
Let us now consider collective excitations:
  a phase fluctuation of the order parameter (``$\pi$-exciton'')
   analogous to the pion in QCD,  
   and an amplitude fluctuation of the order parameter 
  (``$\sigma$-exciton'') 
    analogous to  the $\sigma$-meson in QCD.
   In terms of the auxiliary fields,  the former (the latter)
   corresponds to  $\phi_{\pi}(x)$ ($\phi_{\sigma}(x)$) in the 
    decomposition,
  $\phi(x)= \sigma + \phi_{\sigma}(x) + i \epsilon(x) \phi_{\pi}(x)$. 
   Propagators  of the collective modes within the one-loop
    approximation 
in the strong coupling limit ($\beta=0$)  reads
\begin{eqnarray}
& & \! \! \! \! \! \!  
{D}_{\phi_{\sigma,\pi}}^{-1}(\bp,i \omega_*) \label{eq:D-prop}
 \\
& & \! \! \! \! \! \! = \frac{1}{2}  - \frac{1+\cosh \omega_*}{8} \int_{k}
H(\bk,\bp;\sigma) G(\bk;\sigma) G(\bk+\bp;\sigma) ,\nonumber
\end{eqnarray}
where $H(\bk,\bp;\sigma)
= \sum_{j=1,2} \sin k_j \sin (k_j+p_j) \pm (m_*+\sigma/2)^2 $
 with  the $+$ ($-$) sign corresponding to the $\pi$-exciton (the $\sigma$-exciton). 
 The $O(\beta)$ correction to the above expression can also be obtained.
The actual dispersion relation without the scale transformation of the 
    time variable $\tau$  is derived from
   the pole of the Euclidean propagator, 
   $D_{\phi_{\sigma,\pi}}^{-1}(\bp,i \omega/v_{_F}) = 0$.

\begin{figure}[tb]
\begin{center}
\includegraphics[width=7cm]{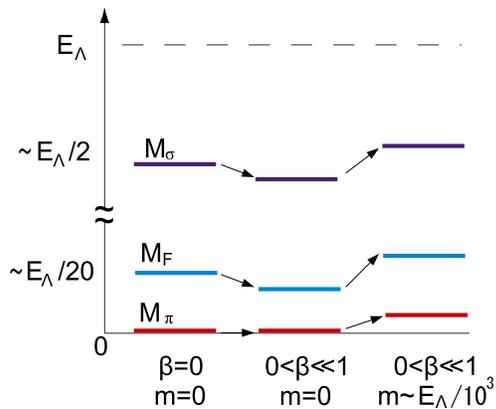}
\end{center}
\caption{Typical mass gaps for the fermion ($M_F$),
 the $\pi$-exciton ($M_{\pi}$) and 
 the $\sigma$-exciton ($M_{\sigma}$).
 The cutoff energy of our effective theory
 reads $E_{_\Lambda} = v_{_F} (\pi/a) \lesssim v_{_F} (\pi/a_{\rm hc}) \sim 10~{\rm eV}$.}
\label{fig:spectrum}
\end{figure}

 In the chiral limit,  $D_{\phi_{\pi}}^{-1}({\bf 0},0) = 0$
 is identical to the gap equation, 
 $\partial F_{\rm eff}(\sigma)/\partial \sigma =0$, so that
 the $\pi$-exciton is indeed a NG boson 
 associated with dynamical breaking of chiral symmetry.
  For the leading order in $m$, the 
  $\pi$-exciton mass, $M_{\pi}=\omega_{\pi}({\bf p}={\bf 0})$, reads, 
\begin{eqnarray}
\label{eq:mpi}
M_\pi \simeq 
2 \sqrt{\frac{m}{M_F^{m=0}}} \left(\frac{v_{_F}}{a}\right) 
\end{eqnarray}
As long as $0 \le m < 2 \ {\rm meV}$ is satisfied for $a \sim a_{\rm hc}$,
$M_{\pi} < M_F $ holds, so that the $\pi$-exciton is the 
 lightest mode in  the system.
The relation, $M_{\pi} \propto \sqrt{m}$, is similar to 
the GOR relation  for the pion
 obtained from current algebra in QCD \cite{gmor,Hatsuda:1994pi}:
 Indeed,  the axial Ward--Takahashi identity 
 for the present system is  
 $\langle (\partial_{\mu} J^{\rm axial}_{\mu}(x) - 2m P(x)) P(y) - 2M(y) \delta_{xy}
  \rangle =0$ 
 with the axial current
 $J^{\rm axial}_{\mu}(x) \equiv \frac{i}{2} \epsilon(x) (V_{\mu}^-(x) - V_{\mu}^+(x))$
 and  the pseudoscalar density $P(x) \equiv i\epsilon(x) M(x)$.  
  Saturating this identity by the 
  $\pi$-exciton  and using Eq.(\ref{eq:mpi}), \cite{Hatsuda:1994pi} we obtain,
  in the leading order of $m$,
 \begin{equation}
(F^{\tau}_{\pi}{M_\pi})^2 = m \sigma , 
\ \ F^{\tau}_{\pi}= {\sigma a^2}(8v_{_F}a)^{-1/2},
\end{equation}
where $\sigma$ takes the value in the chiral limit ($m=0$)
 and  the temporal ``pion decay constant'' $F_{\pi}^{\tau}$ is defined
 by the matrix element, $\langle 0 | J^{\rm axial}_4 | \pi \rangle 
 = 2  F_{\pi}^{\tau} \omega_{\pi}$.
 The dispersion relation for the $\pi$-exciton  is also
 obtained from Eq.(\ref{eq:D-prop}):  
 At low momentum with $\beta=m=0$,  we have
 $ \omega_\pi \simeq v_{\pi}|\bp| $ with the 
 pion velocity $ v_{\pi} = 4.69 v_{_F} = 0.0141$.
As for the mass of the $\sigma$-exciton, we obtain,
$M_{\sigma} \simeq (1.30 - 0.47 \beta)(v_{_F}/a)+22.6m$ by solving 
 $D^{-1}_{\phi_{\sigma}}({\bf 0},iM_{\sigma}/v_{_F}) =0$.
This is comparable to the cutoff energy scale of the present lattice,
 $E_{_{\Lambda}}  \equiv v_{_F}(\pi/a)$.
Thus the result is
not universal unlike the $\pi$-exciton case,
 and analogous to the situation for
the broad $\sigma$-meson in QCD \cite{Hatsuda:1994pi}.
Shown in Fig.\ref{fig:spectrum} is a qualitative summary of the 
 spectra of the fermion and collective excitations obtained in this study.
 
As shown in Refs.\onlinecite{drut_2009}, \onlinecite{giedt_2009} and \onlinecite{drut_2010},
 one may employ a non-compact formulation of the gauge action,
$S_G^{\rm (NC)} = \frac{1}{2g_*^2} 
 \sum_{x^{(4)}} \sum_{j=1,2,3} \left[\theta(x)-\theta(x+\hat{j})\right]^2$
in order to avoid anomalous phase transition from magnetic monopole condensation
\cite{monopole-cond}.
Within the NLO, we find that 
the non-compact results are obtained by the rescaling, e.g.
$\sigma^{\rm (NC)}(\beta) = \sigma^{\rm (C)}(2\beta) + O(\beta^2)$.
Our relations $\sigma^{\rm (NC)}(\beta=0) = \sigma^{\rm (C)}(\beta=0)$
and $\sigma^{\rm (NC)}(\beta) < \sigma^{\rm (C)}(\beta)$ in the strong coupling region
are consistent with the recent lattice Monte Carlo simulations \cite{drut_2010}.

In this Rapid Communication, we performed an analytical  study of
  the monolayer graphene in the strong coupling regime 
  of   U(1) lattice gauge theory.
 An effective action at zero temperature  for the
  Dirac quasiparticles is derived up to
   next-to-leading order of the strong coupling expansion.
 Dynamical breaking of chiral symmetry and associated 
  formation of a chiral gap are found.
 We showed that the $\pi$-exciton (similar to the pion in QCD)
 behaves as a pseudo Nambu--Goldstone
  boson in the strong coupling regime.
   A mass formula for the $\pi$-exciton
  analogous to the Gell-Mann--Oakes--Renner
   relation in QCD  is derived.
If the GOR-type mass formula of the $\pi$-exciton 
can be experimentally confirmed, e.g. through transport phenomena,
it would be a good evidence for the dynamical chiral symmetry breaking in monolayer graphene.

 There are numerous problems to be examined in the future.
 Generalization of our approach with the tadpole improvement \cite{giedt_2009,drut_2010} can be performed. 
 To study the renormalization effect on $v_{_F}$, we need to consider the 
  excitonic fluctuations acting on Dirac quasiparticles.  
 These fluctuations are also important for  thermal phase transition of graphene 
 from insulator to semimetal.
 To study universal low-energy behavior of the graphene,
  it would be useful to construct a
  chiral effective theory for light $\pi$-excitons,  a non-covariant
   analogue of the  chiral perturbation theory in QCD
    \cite{lewtwyler_1994}.
 To be more faithful to the U(4) chiral symmetry at low 
  energies in Eq.(\ref{eq:effaction}),
   we should employ     lattice gauge theory
      with domain-wall or overlap fermions.
  Finally, to study  multilayer graphene where 
  the inter-layer electron hopping depends on how the layers are stacked, 
  it would be important to develop a lattice gauge theory
  preserving the original honeycomb  structure.

\begin{acknowledgments}
 We thank H. Aoki, T. Z. Nakano, Y. Nishida, 
 A. Ohnishi, T. Oka,  S. Sasaki and  N. Yamamoto
  for discussions. This work was supported in part by
 the Grant-in-Aid for 
 Scientific Research, MEXT, Japan (No. 18540253 and No. 2004: 20105003).
Y. A. is supported by Grant-in-Aid for Japan Society for the Promotion of Science
(DC1, No.22.8037).

\end{acknowledgments}



\begin{thebibliography}{99}
\bibitem{chiral-symmetry}
This U(4) symmetry is an approximate one valid only at low energies and is  
perturbed by  symmetry breaking terms if 
 higher order terms in momentum are taken into account.
\bibitem{NG4}
Note here that $2N$-fermi terms with 
 $N \ge 4$ cannot be treated in the extended Stratonovich-Hubbard
  transformation and require different theoretical treatment.
\end{thebibliography}
\end{document}